\documentstyle[11pt,twoside,preprint,aps,epsf]{revtex}

\tightenlines

  \def \bm #1{ \bbox{#1} }
 \newcommand{\be}{\begin{eqnarray}}
 \newcommand{\ee}{\end{eqnarray}}
\begin{document}
%\draft
\thispagestyle{empty}
\begin{center}
{\footnotesize Available at: 
{\tt http://www.ictp.trieste.it/\~{}pub$_-$off}}\hfill IC/2000/1\\
\vspace{1cm}
United Nations Educational Scientific and Cultural Organization\\
and\\
International Atomic Energy Agency\\
\medskip
THE ABDUS SALAM INTERNATIONAL CENTRE FOR THEORETICAL PHYSICS\\
\vspace{1.5cm}
{\bf POSSIBLY TO TEST THE MECHANISM OF ELASTIC
     BACKWARD PROTON-DEUTERON SCATTERING ?}\\
\vspace{1.0cm}
A.Yu.~Illarionov\footnote{E-mail: illar@thsun1.jinr.ru}\\
{\em The Abdus Salam International Centre for Theoretical
Physics, Trieste, Italy}\\
{\em and }\\
{\em Joint Institute for Nuclear Research, Dubna, Russia.}\\[0.2cm]
and\\[0.2cm]
G.I.Lykasov\footnote{E-mail: lykasov@nusun.jinr.ru}\\
{\em Joint Institute for Nuclear Research, Dubna, Russia.}
\end{center}
\vspace{0.5cm}
\centerline{\bf Abstract}
\bigskip
\baselineskip=16pt

 The elastic backward proton-deuteron scattering is analyzed within a
covariant approach based on the invariant expansion of the reaction
amplitude. The relativistic invariant equations for all the polarization
observables are presented. Within the impulse approximation the relation
of the tensor analyzing power $T_{20}$ and the polarization transfer
$\kappa_0$ to $P$-wave components of the deuteron wave function
is found. The comparison of the theoretical calculations with experimental
data is presented. An experimental verification of the reaction 
mechanism is suggested by constructing some combinations of different    
observables.

%\pacs{PACS number(s): 21.45.+v, 21.10.Ky, 21.60.-n, 24.70.+s}
%\hfill{ To appear in \em{Phys.Rev.C}}
\vfill
\begin{center}
MIRAMARE -- TRIESTE\\
\medskip
December 2000\\
\end{center}
\vfill

\newpage

\baselineskip=18pt

As known, the study of polarization phenomena in hadron and hadron-nucleus 
collisions gives more detailed information about dynamics of their 
interaction and the structure of colliding particles. Among the simplest
reactions with hadron probes are processes of forward or backward scattering
of protons off the deuteron. In particular the tensor analyzing power $T_{20}$ 
by backward $pD$ elastic scattering has been measured in Saclay yet fifteen 
years ago \cite{Arv84}. These interesting data yet can't be understood
theoretically especially at the kinetic energy of protons emitted backward
$T_p > 0.6$ GeV. The intensive experimental study of the elastic and 
inelastic $pD$ reaction has been continued in Dubna and Saclay (see for
instance \cite{Azh94,Pun95}) and is also planed to be investigated in the 
nearest future at COSY \cite{Kom95}. All these data can't be described within
the impulse approximation by using the usual deuteron wave function having 
only $S$- and $D$-waves as it is shown in \cite{Ker69}.

In this paper we concentrate our attention on the study of the contribution
of a possible $P$-wave component in the deuteron wave function (DWF) by
using helicity amplitudes formalism to all the polarization observables and
in particular such as the tensor analyzing power $T_{20}$ and deuteron-proton 
polarization transfer $\kappa_0$. This contribution is investigated within the
impulse approximation. We suggest an experimental test of the reaction
mechanism by measuring some combinations of the polarization characteristics.  
       
\begin{figure}[]
 \epsfxsize=5cm \hspace{5cm} \epsfbox{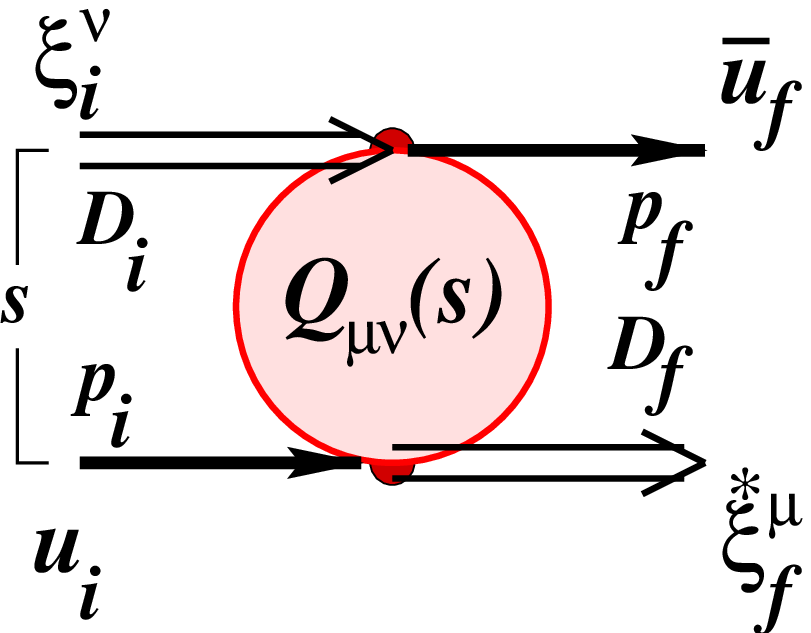}
\caption [A Picture]
{\protect\normalsize
Elastic backward proton-deuteron amplitude.
}
\label{fig:General}
\end{figure}

% \section{General formalism}
% \label{sec:general}
%
  \vspace{0.2cm}
% \subsection{\bf
 $\bullet~~${\bf Invariant expansion of $pD \to Dp$ backward reaction
 amplitude} \\
% \label{subsec:expansion}
%
\noindent
 Let us start from the basic relativistic invariant expansion of elastic
 backward proton-deuteron amplitude using Itzykson-Zuber
 conventions \cite{Itz80}. (see FIG.\ref{fig:General})
%
%\begin{figure}
% \epsfxsize=5cm \hspace{6cm} \epsfbox{pict/pd-dp-Q.eps}
%\caption{Elastic backward proton-deuteron amplitude.}
%\label{fig:General}
%\end{figure}
%

 In general case the relativistic amplitude for the elastic scattering
 of two particles with spins $1$ and $1/2$ has $12$ relativistic
 invariant amplitudes, if all particles are on-mass shell and taking into
 account the $P$- and $T$-invariance. ($3 \times 2 \times 3 \times 2 =
 36$, $P$-invariance results in $18$ functions and $T$-invariance results
 in $12$ functions). The general form for the amplitude of reaction
 $pD\to Dp$ can be written in the following form:
 \begin{equation}
 {\cal M}_{\sigma_f\sigma_i}^{\beta_f\beta_i}(s, t, u) ~=~
 \left[\bar u_{\sigma_f}(p_f) ~{\cal Q}^{\mu\nu}(s, t, u)~
            u_{\sigma_i}(p_i)\right]
 \xi^{*(\beta_f)}_\mu(D_f) ~ \xi^{~(\beta_i)}_\nu(D_i),
 \label{IA1}
 \end{equation}
 where $u_{\sigma_i}(p_i) \equiv u_i$ and
 $\bar u_{\sigma_f}(p_f) \equiv \bar u_f$ are the spinors of the initial and
final nucleons with spin projections $\sigma_i$ and $\sigma_f$ respectively;
$\xi_\mu(D)$ is the polarization vectors of deuterons; $s,t,u$ are invariant
Mandelstam's variables
 \begin{equation}
 s = (D_i + p_i)^2~~;~~t = (D_i - D_f)^2~~;~~u = (D_i - p_f)^2 = \bar s~~,
 \label{mand}
 \end{equation}

 For the backward $pD \to Dp$ scattering the amplitude (\ref{IA1})
 depends only on the one kinematical variable which is chosen usually as
 $s$, e.g., square of the initial energy in the c.m.s.
 The amplitude ${\cal Q}_{\mu\nu}$ for this process contents four
 amplitudes and can be written in the form:
 \begin{equation}
 {\cal Q}_{\mu\nu}(s) ~=~
  {\cal Q}_0(s) \left(-g_{\mu\nu} + q_\mu q_\nu\right) +
  {\cal Q}_1(s) q_\mu q_\nu +
  {\cal Q}_2(s) q_{\{\mu}\gamma_{\nu\}} +
 i{\cal Q}_3(s) \gamma_5\varepsilon_{\mu\nu\rho\sigma}\gamma^\rho q^\sigma~,
 \label{IA2}
 \end{equation}
 where we introduce the unit 4-vector $q = Q/\sqrt{Q^2},~Q = (D_i +
 D_f)/2$.

 \vspace{0.2cm}
 $\bullet~~$ {\bf Helicity amplitudes}\\
% \label{subsec:helicity}
 To calculate the observables, differential cross sections and polarization
characteristics, it would be very helpful to construct the helicity
amplitudes of the considered process $pD \to Dp$. Let us introduce initial
(final) proton helicities $\mu_{i,f} = \pm 1/2$ and the initial (final)
deuteron helicities $\lambda_{i,f} = \pm 1, 0$.
The number of independent helicity amplitudes is the same as the one for
corresponding amplitudes incoming to ${\cal Q}^{\mu\nu}(s)$ (\ref{IA2}) and
equal to four. They can be chosen as the following
\begin{eqnarray}
&&\Phi_{^1_3} = {\cal M}_{+-}^{\pm\mp} = -{\cal M}_{-+}^{\mp\pm}~;~~
  \Phi_2 = {\cal M}_{+-}^{00} = -{\cal M}_{-+}^{00}~; \nonumber \\
&&\Phi_{4} = {\cal M}_{++}^{+0} = -{\cal M}_{++}^{0+} =
  {\cal M}_{--}^{0-} = -{\cal M}_{--}^{-0}~,
\label{HA}
\end{eqnarray}
They are related to each other by using the symmetry properties: \\
$\bullet~~$ Parity
\begin{equation}
{\cal M}_{-\mu_f-\mu_i}^{-\lambda_f-\lambda_i} = (-1)^{2(\mu_i-\lambda_i)}
{\cal M}_{\mu_f\mu_i}^{\lambda_f\lambda_i}~;
\label{H8.1}
\end{equation}
$\bullet~~$ Time - reversal
\begin{equation}
{\cal M}_{\mu_f\mu_i}^{\lambda_f\lambda_i} = (-1)^{2(\mu_i-\lambda_i)}
{\cal M}_{\mu_i\mu_f}^{\lambda_i\lambda_f}~.
\label{H8.2}
\end{equation}
The Eqs.(\ref{H8.1},\ref{H8.2}) result in the following relation: 
${\cal M}_{-\mu_f-\mu_i}^{-\lambda_f-\lambda_i} =
 {\cal M}_{\mu_i\mu_f}^{\lambda_i\lambda_f}~.$

 Using the expansion (\ref{IA2}) for ${\cal Q}^{\mu\nu}(s)$ one can
 relate the relativistic invariants ${\cal Q}_i$ to the corresponding
 helicity amplitudes $\Phi_i$:
 \begin{mathletters}
 \label{Phi}
 \begin{eqnarray}
 &&\Phi_{^1_3} = {\varepsilon \over m}{\cal Q}_0 \pm {\cal Q}_3~;
 \label{Phi13} \\
 &&\Phi_2 = -{\varepsilon \over m}{\cal Q}_0 - {p^2\over M^2}
 \left({\varepsilon \over m}[{\cal Q}_0 - {\cal Q}_1] - 2{\cal Q}_2\right)~;
 \label{Phi2} \\
 &&\Phi_{4} = -\sqrt{2}{p^2 \over Mm}{\cal Q}_2 -
 \sqrt{2}{\varepsilon\varepsilon_D \over Mm}{\cal Q}_3~.
 \label{Phi4}
 \end{eqnarray}
 \end{mathletters}
 And these helicity amplitudes can be related to the corresponding
 Pauli's amplitudes $g_i$:
 \begin{equation}
 \Phi_{^1_3} = g_1 \mp g_4~;~~\Phi_2 = -g_2~;~~\Phi_4 = \sqrt{2}g_3~.
 \label{HP}
 \end{equation}

 \vspace{0.0cm}
 $\bullet~~$ {\bf Polarization observables} \\
 % \label{subsec:observables}
 Having the helicity amplitudes given by Eq.(\ref{HA}) one may define
 various polarization characteristics for the discussed process. 
 Applying the notations used in Refs. \cite{Bou80,Gha91} we define the set
 of all the possible polarization observables as the following:
 \begin{equation}
 \left( \alpha; \mu | \beta; \nu \right) =
  {Tr\left[\sigma_\alpha {\cal O}_\mu {\cal M}^+
           \sigma_\beta {\cal O}_\nu {\cal M} \right] \over
   Tr\left[{\cal M}^+ {\cal M} \right]}~,
 \label{Ob0}
 \end{equation}
 with a normalization $\left( 0; 0 | 0; 0 \right) = 1$. The subscripts
 $\alpha$ and $\mu$ ($\beta$ and $\nu$) refer to the polarization
 characteristics of the initial (final) proton and deuteron respectively;
 $\sigma_\alpha$ is the Pauli matrix, and ${\cal O}_\mu$ stands for a set
 of $3 \times 3$ operators defining the deuteron polarization. The quantity
 $\Sigma = Tr\left[{\cal M}^+ {\cal M} \right]$
 \begin{equation}
 \Sigma = \sum_{\text{\scriptsize all} {\hspace{0.05cm}} \mu,\lambda}
 |{\cal M}_{\mu_f\mu_i}^{\lambda_f\lambda_i}(W)|^2 =
 2(|\Phi_1|^2 + |\Phi_2|^2 + |\Phi_3|^2 + 2|\Phi_4|^2)
 \label{CS1}
 \end{equation}
 is related to the unpolarized differential cross section as
 \begin{equation}
 {d\sigma \over d\Omega} =
 {1\over6}\left({m \over 4\pi\sqrt{s}}\right)^2 \cdot \Sigma =
 \sigma_0 \cdot \Sigma~,
 \label{CS} 
 \end{equation}
 Using the time-reversal invariance one can get the relation:
 $\left( \alpha; \mu | \beta; \nu \right)_{\pi} =
 \left( \beta; \nu | \alpha; \mu\right)_{\pi}$.
 Another relation as the consequence from the parity invariance is
 $\left( \alpha; \mu | \beta; \nu \right) = 0$ if $n_L + n_S$ is odd,
 where $n_{L,S}$ are the numbers of the indexes $L$ or $S$ appearing
 in the symbols $\alpha, \mu, \beta, \nu$.

 As mentioned in Ref.\cite{Lad97}, one of the goals of the future
 experiments is a direct reconstruction of the complex amplitudes (\ref{HA}).
 A overfull set of polarization observables for complete measurement has
 been proposed in Refs.\cite{Lad97,Lad96}. In terms of the helicity
 amplitudes this overfull set can be written as the following:
 {\small
 \begin{mathletters}
 \label{observ}
 \begin{eqnarray}
 &&\left(0; NN | 0; 0 \right) = \left(0; 0 | 0; NN \right) =
 -\left[|\Phi_1|^2 - 2|\Phi_2|^2 + |\Phi_3|^2 - |\Phi_4|^2\right] \cdot
 \Sigma^{-1} = A_{yy} = -T_{20}/\sqrt{2}~;
 \label{T20} \\
 &&\left(0; N | 0; N \right) = \left(0; S | 0; S \right) =
 -2\left[\text{Re}\left(\Phi_1 + \Phi_3\right)\Phi_2^* -
 |\Phi_4|^2\right] \cdot \Sigma^{-1} = D_{N} = D_{S} = A_{y}~;
 \label{Ay} \\
 &&\left(0; L | 0; L\right) =
 2\left[|\Phi_1|^2 + |\Phi_3|^2\right] \cdot \Sigma^{-1} = D_{L}~;
 \label{DL} \\
 &&\left(0; LL | 0; LL\right) = -2 \left(0; NN | 0; LL \right) =
 2\left[|\Phi_1|^2 + 4|\Phi_2|^2 + |\Phi_3|^2 - 4|\Phi_4|^2\right] \cdot
 \Sigma^{-1} = D_{LL}~;
 \label{DLL} \\
 &&\left(0; NN | 0; NN\right) =
 \left[|\Phi_1|^2/2 + 9\text{Re}(\Phi_1\Phi_3^*) + |\Phi_3|^2/2 +
 2|\Phi_2|^2 - 2|\Phi_4|^2\right] \cdot \Sigma^{-1} = D_{NN}~;
 \label{DNN} \\
 &&\left(0; NN | 0; SS\right) =
 \left[|\Phi_1|^2/2 - 9\text{Re}(\Phi_1\Phi_3^*) + |\Phi_3|^2/2 +
 2|\Phi_2|^2 - 2|\Phi_4|^2\right] \cdot \Sigma^{-1}~;
 \label{NN-SS} \\
 &&\left(0; SN | 0; SN\right) =
 \left[ \left(0; NN | 0; NN \right) - \left(0; NN | 0; SS \right) \right] / 2 =
 9\text{Re}(\Phi_1\Phi_3^*) \cdot \Sigma^{-1} = D_{SN}~;
 \label{DSN} \\
 &&\left(0; LN | 0; LN\right) = \left(0; LS | 0; LS \right) =
 -\left( 9/2 \right) \left[\text{Re}\left(\Phi_1 + \Phi_3\right)\Phi_2^* +
 |\Phi_4|^2\right] \cdot \Sigma^{-1} = D_{LN} = D_{LS}~;
 \label{DLS} \\
 &&\left(0; N | 0; LS\right) = \left(0; LN | 0; S \right) =
 -\left(0; S | 0; LN \right) = -\left(0; LS | 0; N \right) =
 3\text{Im}\left[\left(\Phi_1 + \Phi_3\right)\Phi_2^*\right] \cdot
 \Sigma^{-1}~;
 \label{N-LS} \\
 &&\left(L; L | 0; 0\right) = \left(0; 0 | L; L \right) =
 -2\left[|\Phi_1|^2 - |\Phi_3|^2 + |\Phi_4|^2\right] \cdot \Sigma^{-1} =
 2A_l~;
 \label{LL-00} \\
 &&\left(N; N | 0; 0\right) = \left(0; 0 | N; N \right) =
 \left(0; 0 | S; S \right) = \left(S; S | 0; 0 \right) =
 2\sqrt{2}\text{Re}\left[\left(\Phi_1 - \Phi_2\right)\Phi_4^*\right] \cdot
 \Sigma^{-1} = 2A_t~;
 \label{At} \\
 &&\left(N; LS | 0; 0\right) = \left(0; 0 | N; LS \right) =
 -3\sqrt{2}\text{Im}\left[\left(\Phi_1 + \Phi_2\right)\Phi_4^*\right] \cdot
 \Sigma^{-1}~;
 \label{NLS-00} \\
 &&\left(0; L | L; 0\right) = \left(L; 0 | 0; L \right) =
 -2\left[|\Phi_1|^2 - |\Phi_3|^2 - |\Phi_4|^2\right] \cdot \Sigma^{-1}
 = -(4/3)\kappa_l~;
 \label{kl} \\
 &&\left(0; N | N; 0\right) = \left(N; 0 | 0; N \right) =
 2\sqrt{2}\text{Re}\left[\left(\Phi_3 - \Phi_2\right)\Phi_4^*\right] \cdot
 \Sigma^{-1} = (4/3)\kappa_t = (2/3)\kappa_0~;
 \label{k0} \\
 &&\left(0; LS | N; 0\right) = \left(N; 0 | 0; LS \right) =
 -3\sqrt{2}\text{Im}\left[\left(\Phi_3 + \Phi_2\right)\Phi_4^*\right] \cdot
 \Sigma^{-1} = 3A_4~;
 \label{A4} \\
 &&\left(L; 0 | L; 0\right) =
 \left[|\Phi_1 + \Phi_3|^2 + 2|\Phi_2|^2 - 2|\Phi_4|^2\right] \cdot
 \Sigma^{-1} = P_{L}~;
 \label{PL}\\
 &&\left(N; 0 | N; 0\right) = \left(S; 0 | S; 0 \right) =
 2\left[2\text{Re}(\Phi_1\Phi_3^*) + |\Phi_2|^2\right] \cdot \Sigma^{-1}
 = P_{N} = P_{S}~.
 \label{PS}
 \end{eqnarray}
 \end{mathletters}
 }
We use a righthand coordinate system, defined in accordance with
Madison convention \cite{madison}. This system is specified by a
set of three orthogonal vectors $\bm L$, $\bm N$ and  $\bm S$,
where $\bm L$ is the unit vector along the momenta  of the incident
particle, $\bm N$ is taken to be orthogonal to $\bm L$,
$\bm S = \bm N \times \bm L$.

 Since the process is described by using four complex amplitudes, one 
 needs to measure at least seven independent observables. The magnitudes
 of amplitudes $|\Phi_i|$ can be extracted from the
 cross section $\Sigma$, tensor analyzing power $T_{20}$,
 tensor-tensor spin transfer coefficient sum:
 $D = [(0; NN | 0; NN) + (0; NN | 0, SS)]$ and the spin correlation
 parameter $\kappa_l$.
 It has to be noted, since the observables have forms as bilinear
 combinations of the amplitudes, the finding of the common phase is
 impossible. For simplicity we put the phase of the
 amplitude $\Phi_3$ equal to zero: $\varphi_3 = 0 ~\to~
 {\text{Im}}(\Phi_3) = 0,~{\text{Re}}(\Phi_3) = \sqrt{\Phi_3^2} =
 |\Phi_3|$.
 Then, the other phases can be obtained from the three
 observables: the spin transfer coefficient from the deuteron to proton,
 $\kappa_0$, and the spin correlation parameters $(N; N | 0; 0)$ and
 $(N; LS | 0; 0)$. These observables are mostly realistic to be
 measured at the moment with the existing experimental techniques
 \cite{Lad96}.

\begin{figure}[]
 \epsfxsize=5cm \hspace{5cm} \epsfbox{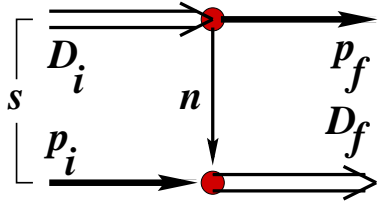}
\caption [A Picture]
{\protect\normalsize
The one-nucleon exchange diagram.
}
\label{fig:ONE}
\end{figure}

 \vspace{0.2cm}
 $\bullet~~$ {\bf The one-nucleon exchange mechanism (ONE)} \\
% \label{ONE}
 Let us consider our reaction within the framework of the impulse
 approximation, FIG.\ref{fig:ONE}.
%
%\begin{figure}
% \epsfxsize=5cm \hspace{6cm} \epsfbox{pict/pd-dp.eps}
%\caption{The one-nucleon exchange diagram.}
%\label{fig:ONE}
%\end{figure}
%
 In ONE model the amplitude of the $ pD \to Dp$ backward reaction has a
 very simple form \cite{Ker69}:
 \begin{equation}
 {\cal Q}_{\mu\nu}^N = \Gamma_\nu {\widehat n - m \over m^2 - u}
                      \bar\Gamma_\mu~,
 \label{ONE1}
 \end{equation}
 where $\Gamma_\nu (\bar\Gamma_\mu = \gamma_0\Gamma^+_\mu\gamma_0)$ is a
 deuteron vertex with one off-shell nucleon and can be written with four
 form factors parameterization exactly coinciding with the one used, for
 instance, by Gross \cite{Gro79,Gro92,Gro82} or Keister and Tjon
 \cite{Kei82,Kei81}:
 \begin{equation}
 \Psi_\nu = {\Gamma_\nu(D,q) \over m^2 - n^2 - i0} =
 \varphi_1(u)\gamma_\nu + \varphi_2(u){n_\nu\over m} +
 \left(\varphi_3(u)\gamma_\nu + \varphi_4(u){n_\nu \over m}\right)
 {\widehat n + m\over m}~.
 \label{Psi}
 \end{equation}
 The form factors $\varphi_i(u)$ are invariant scalar functions depending on
 the invariant $n^2 = u$ may be computed in any reference frame. 
 To connect this relativistic invariant formalism with the
 non-relativistic one we also express $\varphi_i$ in the
 deuteron rest frame in terms of partial amplitudes, namely in the
 $\rho$-spin classification,  the two large components of the DWF
 $U = ^3{\cal S}^{++}_1$ and $W = ^3{\cal D}^{++}_1$, and the small
 components $V_s = ^1{\cal P}^{+-}_1$ and $V_t = ^3{\cal P}^{+-}_1$ 
 as like as in \cite{Gro79}.

 By substituting Eq.(\ref{Psi}) into Eq.(\ref{ONE1}) and making use of
 the
 identities $n = D_i - p_f = D_f - p_i,~ n^2 = u \leq (M - m)^2$, after
 computing the quantities (\ref{IA2}), one can find the forms of the
 helicity amplitudes (\ref{Phi}) within the ONE model in terms of this
 positive- and negative-energy wave functions:
 \begin{mathletters}
 \label{ONE:Phi}
 \begin{eqnarray}
 &&\Phi_1^N(W) = 0~;
 \label{ONE:Phi1}\\
 &&\Phi_2^N(W) = -2\pi^2\left(m^2 - u\right)\left[{\varepsilon_D \over M}
  \left(U + \sqrt2W\right) - 2\sqrt3{p\over m}V_s\right]
  \left(U + \sqrt2W\right) - 6\pi^2M\varepsilon_D V_s^2~;
 \label{ONE:Phi2}\\
 &&\Phi_3^N(W) = 2\pi^2\left(m^2 - u\right)\left[{\varepsilon_D \over M}
  \left(\sqrt2U - W\right) - 2\sqrt3{p\over m}V_t\right]
  \left(\sqrt2U - W\right) + 6\pi^2M\varepsilon_D V_t^2~;
 \label{ONE:Phi3}\\
 &&\Phi_4^N(W) = 2\pi^2\left(m^2 - u\right)\left[{\varepsilon_D \over M}
  \left(\sqrt2U - W\right)\left(U + \sqrt2W\right)\right.
 \nonumber \\&&\hspace{3.5cm}\left.
 - \sqrt3{p\over m}
  \left\{\left(\sqrt2U - W\right)V_s +
  \left(U + \sqrt2W\right)V_t\right\}\right]
%\nonumber \\&&\hspace{1cm}
  +~6\pi^2M\varepsilon_D V_sV_t.
 \label{ONE:Phi4}
 \end{eqnarray}
 \end{mathletters}
 where $P_{\text{lab}}$ is the final proton momentum.
 Firstly, one can see, that all the $\Phi_i^N(W)$ amplitudes are real,
 e.g., all the T-odd polarization correlations are equal to zero within
 this approximation.
 For example, $(N; LS| 0; 0) = 0$.
 Secondly, within the ONE approximation the helicity amplitude
 $\Phi_1^N(W)$ is vanished because the spin-down proton in the incident
 channel cannot result in the spin-down deuteron in the final channel 
 due to the lack of the spin non-flip of the proton. This leads to 
           $(0; NN| 0; NN) = (0; NN| 0; SS)$.
 This consequence of the ONE mechanism can be verified experimentally
 by measuring and combining the different observables given by
 Eqs.(\ref{observ}). For example, combine the Eq.(\ref{T20}) and
 Eq.(\ref{kl}) one can find the helicity amplitude $\Phi_1(W)$:
$
 |\Phi_1(W)|^2 = (1 + T_{20}/\sqrt{2} + 2\kappa_l) \cdot \Sigma/6.
$

 And finally, the following relation between amplitudes:
 \begin{equation}
 \Delta^N \equiv \Phi_2^N\Phi_3^N + (\Phi_4^N)^2 =
 -12\pi^4 {M^3 E_{\text{lab}}^2 (2E_{\text{lab}} - M) \over m^2}
  \left[ \left(\sqrt{2}U - W\right)V_s -
  \left(U + \sqrt{2}W\right)V_t\right]^2~
 \label{Delta}
 \end{equation}
 has a purely P-wave dependence. We have for a ``Magic Circle'' in the
 $\kappa_0$-$T_{20}$ plane \cite{Kue93} the following equation:
 \begin{equation}
 {(\kappa_0^N)^2 \over 9/8} + {(T_{20}^N + 1/(2\sqrt{2}))^2 \over 9/8} =
  1 - \left( 4{ {\Delta^N} \over \Sigma^N} \right)^2~.
 \label{circle}
 \end{equation}
 Using general formulas for the polarization observables in terms of the
 $\Phi_i^N(W)$ helicity amplitudes, one can calculate all observables in
 terms
 of positive- and negative-energy wave functions, $U,W$ and $V_s, V_t$
 respectively. The contribution of the positive-energy wave $U, W$ to the
 observables is reffered to as the non-relativistic result. The parts
 containing the negative-energy waves $V_s, V_t$ are of a purely
 relativistic
 origin and consequently they manifest genuine relativistic correction
 effects. Additionally there is another source for the relativistic
 corrections, namely so-called Lorenz boost effects coming from the
 transformation of the DWF from the c.m.s. to the deuteron rest frame
\cite{Kap98}.

 In terms of positive-energy waves $U$ and $W$ only the helicity
 amplitudes
 have a well-known non-relativistic form. For this simple case there is
 the following relation: 
        $\Delta^N = \Phi_2^N \Phi_3^N  + (\Phi_4^N)^2 = 0$.
 And the Lorenz boost effects do not contribute to the polarization
 observables.

\begin{figure}[]
 \epsfxsize=8cm \epsfysize=8cm \hspace{3.5cm} \epsfbox{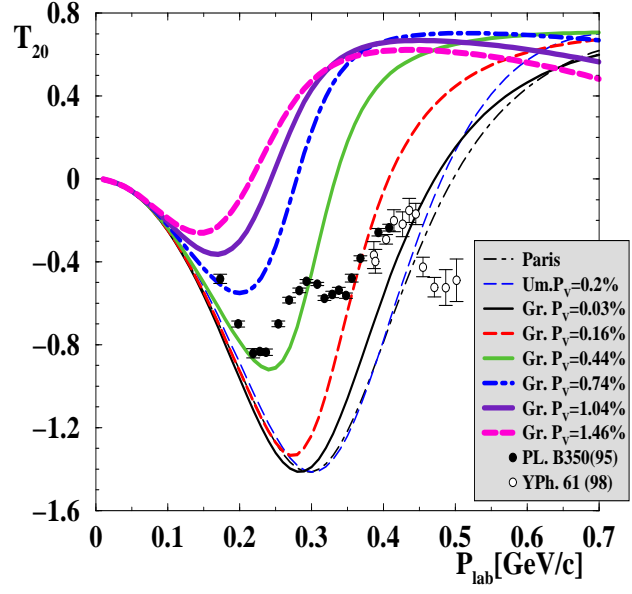}
\caption [A Picture]
{\protect\normalsize
Tensor analyzing power $T_{20}$.
}
\label{fig:T20}
\end{figure}

\begin{figure}[]
\epsfxsize=8cm \epsfysize=8cm \hspace{3.5cm} \epsfbox{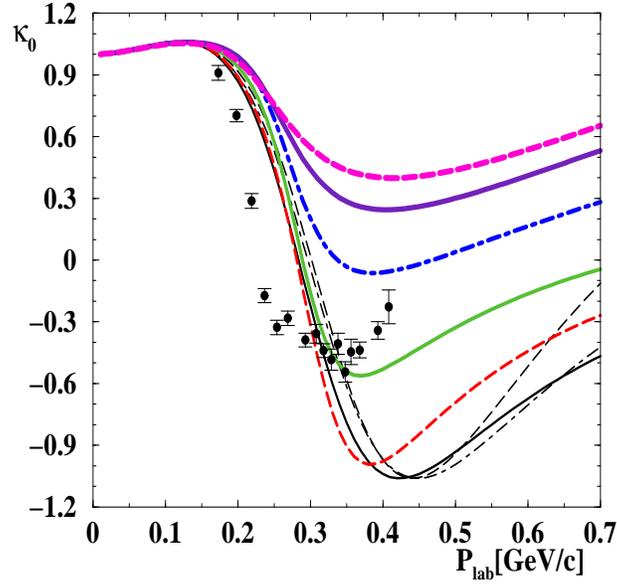}
\caption [A Picture]
{\protect\normalsize
Polarization transfer coefficient $\kappa_0$.
}
\label{fig:k0}
\end{figure}

 \vspace{0.5cm}
 $\bullet~~$ {\bf Results and Discussions} \\
% \label{sec:results}
Let us present the calculation results for the deuteron tensor analyzing 
power $T_{20}$, the polarization transfer $\kappa_0$ and their link given 
by Eq.(\ref{circle}) obtained within the relativistic impulse
approximation. In FIG's.(\ref{fig:T20},\ref{fig:k0}) $T_{20}$ and $\kappa_0$
for different kinds of the DWF are presented. It can be seen from these
figures the inclusion of the $P$-wave to the DWF according to \cite{Gro92}
changes the form of $T_{20}$ at $P_{lab} > 0.2$ GeV/c. The shape of these
observables is changed towards the experimental data by increasing the 
probability of $P$-wave $P_V$ in the DWF. The form of the polarization 
transfer $\kappa_0$ is closed to the experimental data at
$P_V = 0.4\% - 0.5\%$. Although the description of the experimental data
about $T_{20}$ and $\kappa_0$ isn't satisfactory even by inclusion of the
$P$-wave to the DWF nevertheless the $P$-wave contribution improves 
the description of data and shows a big sensitivity of the polarization
observables presented in FIG's.(\ref{fig:T20},\ref{fig:k0}) to this effect. 

\begin{figure}[]
 \epsfxsize=8cm \epsfysize=7cm \hspace{3.5cm} \epsfbox{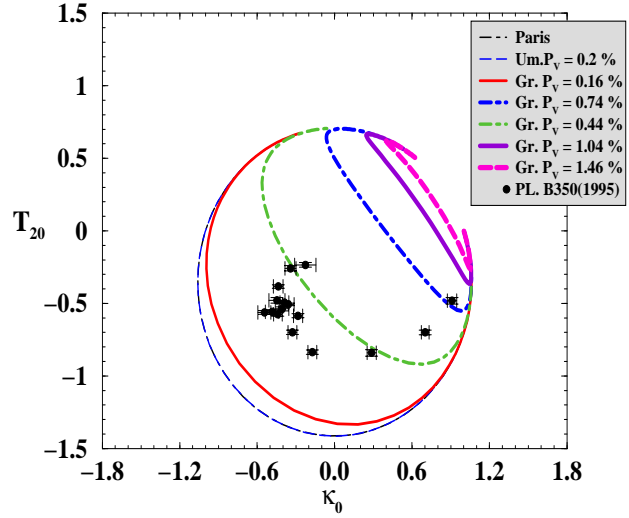}
\caption [A Picture]
{\protect\normalsize
``Magic Circle'' in the $\kappa_0$--$T_{20}$ plane.
}
\label{fig:T20-k0}
\end{figure}

The link between $T_{20}$ and $\kappa_0$ given by Eq.(\ref{circle})
is presented in FIG.\ref{fig:T20-k0}. The big sensitivity of this relation
to the contribution of the $P$-wave probability $P_V$ is also seen from this
figure. There isn't also a satisfactory description of the experimental 
data nevertheless the shape of the "Magic Circle" which is right for 
the conventional DWF is deformed towards the experimental data.

In principle, there is some analogy between the effects of 
the deuteron $P$-wave and secondary interactions contributing to 
the discussed observables for elastic and inelastic backward $pD$
reactions \cite{Kon77,Kon81} and \cite{Lyk}. The contribution of secondary
interactions, in particular the triangle graphs with a pion in
intermediate state, results in an improvement of the description 
of discussed experimental data on observables for the deuteron striping
reaction $Dp \to pX$ \cite{Lyk}.

%
%\begin{figure}
% \epsfxsize=8cm \epsfysize=8cm \hspace{4.5cm}\epsfbox{eps/T20.eps} 
%\caption{Tensor analyzing power.} %\label{fig:T20}
%\end{figure} 
% 
%\begin{figure} 
% \epsfxsize=8cm \epsfysize=8cm\hspace{4.5cm} \epsfbox{eps/k0.eps} 
%\caption{Polarization transfer coefficient.} 
%\label{fig:k0} 
%\end{figure} 
% 
%\begin{figure} 
%\epsfxsize=8cm \epsfysize=7cm \hspace{4.5cm} \epsfbox{eps/T20-k0.eps}
%\caption{``Magic Circle'' in the $\kappa_0$--$T_{20}$ plane.}
%\label{fig:T20-k0} %\end{figure} %\vspace{1.cm} %

 The consequence of the ONE mechanism can be verified experimentally
 by measuring and combining the different observables given by
 Eqs,(\ref{observ}). For example, one can combine the Eq.(\ref{T20}) and
 Eq.(\ref{kl}) in order to find the helicity amplitude $\Phi_1^N(W)$.
 At least, one can find experimentally the kinematical region where 
 $\Phi_1^N(W) = 0$ and the "Magic Circle" Eq.(\ref{circle}) can
 be applicable to find some information about the $P$-wave contribution
 to the DWF.        

 \vspace{0.5cm}
 $~\bullet~$ {\bf Conclusions} \\
 The performed analysis has shown the following. The discussed
 polarization observables $T_{20}$ and $\kappa_0$ are very sensitive 
 to a possible contribution of $P$-wave to the relativistic DWF.
 There is some analogy between the inclusion of $P$-wave to the 
 DWF and effect of the secondary interactions which are some
 corrections to the ONE graph.
 One can propose a verification of the reaction mechanism for the
 elastic backward $pD$ scattering from the measuring of the 
 polarization observable like as $\left(0; SN | 0; SN\right)$ given by
 Eq.(\ref{DSN}), which have to be equal zero within the relativistic
 ONE approximation as it is seen from Eq.(\ref{ONE:Phi1}). Any way,
 combining the another polarization observables which are more available
 for the measurement one can find experimentally whether the helicity
 amplitude $\Phi_1^N(W)$ is equal zero or not at some kinematical region. 
 Therefore one can verify experimentally the validity of the relativistic
 invariant impulse approximation. At least, one can find some kinematical
 region where it is valid more less and extract some information
 about the $P$-wave contribution to the DWF.

%\subsection*{Acknowledgments}

\newpage

%%%%%%%%%%%%%%%%%%%%%%%%%%%%% Figure 1 - Caption

%%%%%%%%%%%%%%%%%%%%%%%%%%%%% Figure 2 - Caption

\newpage

%%%%%%%%%%%%%%%%%%%%%%%%%%%%% Figure 3 - Caption

%%%%%%%%%%%%%%%%%%%%%%%%%%%%% Figure 4 - Caption

\newpage

%%%%%%%%%%%%%%%%%%%%%%%%%%%%% Figure 5 - Caption

\end{document}